\documentclass[prl,twocolumn,showpacs,amsmath,superscriptaddress,psfig,amssymb]{revtex4}

\usepackage{graphicx}
\usepackage{dcolumn}
\usepackage{bm}

\begin{document}


\title{Quantum phase transition in the heavy-fermion compound YbIr$_2$Si$_2$}

\author{H. Q. Yuan}
 \email{yuan@mrl.uiuc.edu} \affiliation {Max Planck Institute for
Chemical Physics of Solids, N\"{o}thnitzer Stra\ss e 40, 01187
Dresden, Germany} \affiliation {Department of Physics, University
of Illinois at Urbana-Champaign, 1110 West Green Street, Urbana,
IL 61801}

\author{M. Nicklas}
\affiliation {Max Planck Institute for Chemical Physics of Solids,
N\"{o}thnitzer Stra\ss e 40, 01187 Dresden, Germany}

\author{Z. Hossain}
\affiliation{Max Planck Institute for Chemical Physics of Solids,
N\"{o}thnitzer Stra\ss e 40, 01187 Dresden, Germany}
\affiliation{Department of Physics, Indian Institute of
Technology, Kanpur-208016, India}

\author{C. Geibel}
\author{F. Steglich}
\affiliation {Max Planck Institute for Chemical Physics of Solids,
N\"{o}thnitzer Stra\ss e 40, 01187 Dresden, Germany}

\date{\today}

\begin{abstract}

We investigate the pressure-temperature phase diagram of
YbIr$_2$Si$_2$ by measuring the electrical resistivity $\rho(T)$.
In contrast to the widely investigated YbRh$_2$Si$_2$,
YbIr$_2$Si$_2$ is a paramagnetic metal below $p_c\simeq 8$ GPa.
Interestingly, a first-order, presumably ferromagnetic, transition
develops at $p_c$. Similar magnetic properties were also observed
in YbRh$_2$Si$_2$ and YbCu$_2$Si$_2$ at sufficiently high
pressures, suggesting a uniform picture for these Yb compounds.
The ground state of YbIr$_2$Si$_2$ under pressure can be described
by Landau Fermi-liquid (LFL) theory, in agreement with the nearly
ferromagnetic Fermi-liquid (NFFL) model. Moreover, evidence of a
weak valence transition, characterized by a jump of the
Kadowaki-Woods (KW) ratio as well as an enhancement of the
residual resistivity $\rho_0$ and of the
quasiparticle-quasiparticle scattering cross section, is observed
around 6 GPa.

\end{abstract}

\pacs{71.27.+a, 71.10.Hf, 73.43.Nq}
\maketitle

The study of quantum critical phenomena has attracted considerable
attention because of the fascinating physical properties caused by
quantum fluctuations. In the Ce-based heavy fermion (HF) systems,
unconventional superconductivity, most likely paired via
antiferromagnetic (AFM) spin fluctuations, has been widely
observed around a spin-density-wave (SDW) type quantum critical
point (QCP) \cite{Thalmeier 2004}. On the other hand,
YbRh$_2$Si$_2$ has been established as a model system to study
quantum physics at a ``local" QCP \cite{Custers 2003}, around
which no superconductivity has yet been observed at $T>10$ mK.
Recent efforts have been largely concentrated on  weakly
first-order quantum phase transitions (QPT), e.g., the
ferromagnetic (FM) transition in MnSi \cite{Thessieu 1995,
Pfleiderer 1997}, the metamagnetic transition in Sr$_3$Ru$_2$O$_7$
\cite{Grigera 2001}, and the valence transition in
CeCu$_2$(Si$_{1-x}$Ge$_x$)$_2$ \cite{Yuan 2003}.

The HF compound YbRh$_2$Si$_2$ undergoes an AFM transition at
$T_N=70$ mK \cite{Trovarelli 2000}. A small magnetic field or a
slight expansion of the unit cell by substituting Si with Ge can
eventually suppress the AFM order at a QCP \cite{Custers 2003,
Gegenwart 2002}, at which the conventional LFL theory breaks down
(see the inset of Fig. 1). As tuning away from the QCP, LFL
behavior immediately recovers at the lowest temperature. On the
other hand, the weak AFM transition in YbRh$_2$Si$_2$ is
stabilized by applying pressure \cite{Trovarelli 2000, Plessel
2003, Knebel 2005}. In particular, the magnetic phase undergoes a
first-order transition from a small-moment state (AFM-type) to a
large-moment state around 10 GPa \cite{Plessel 2003}. Furthermore,
it was argued that in YbRh$_2$Si$_2$ FM quantum critical
fluctuations dominate over a wide range in the phase diagram
except for the close vicinity of the AFM QCP \cite{Ishida 2002,
Gegenwart 2005}. In order to better understand the nature of the
``local" QCP, YbIr$_2$Si$_2$, a sister compound of YbRh$_2$Si$_2$,
was recently synthesized by Hossain {\it et al.} \cite{Hossain
2005}. YbIr$_2$Si$_2$ (I-type) is a moderate HF compound with a
paramagnetic ground state at zero pressure and, therefore, was
expected to cross a magnetic QCP by applying a small pressure of
$2-3$ GPa to compress the unit cell volume of YbIr$_2$Si$_2$ to
that of YbRh$_2$Si$_2$ at $p=0$ \cite{Hossain 2005}.

\begin{figure}
\centering
\includegraphics[width=0.8\columnwidth]{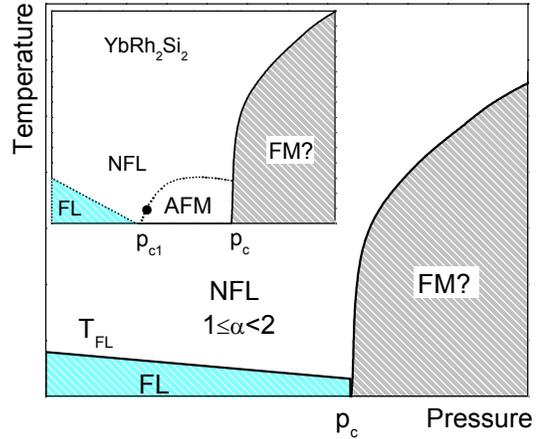}
\caption{(Color online) Schematic phase diagram for the
Yb-compounds other than YbRh$_2$Si$_2$. A large-moment order,
presumably FM-type, develops at a critical pressure $p_c$. The
ground-state properties in both the high-{\it p} phase and the
paramagnetic state can be described by LFL theory. Uniquely, in
YbRh$_2$Si$_2$ (see the inset, the dot marks its location at
$p=0$) AFM phase with weak magnetic moments exists below $p_c$ and
vanishes at a QCP ($p_{c1}$) where the LFL theory is violated.}
\end{figure}

In this letter, we present the first high-pressure study for
YbIr$_2$Si$_2$. Surprisingly, YbIr$_2$Si$_2$ exhibits properties
distinct from YbRh$_2$Si$_2$ and shows no evidence for the
existence of an AFM QCP at low pressures (see Fig. 1). However, a
similar (likely ferro-) magnetic transition is found at high
pressures ($p>p_c\simeq8$ GPa). LFL behavior, characterized by
$\Delta \rho \sim T^2$, survives in the lowest temperature region.
These findings are consistent with the NFFL model \cite{Pfleiderer
1997} and suggest that a first-order FM QPT likely exists in the
pressurized Yb-compounds. Furthermore, evidence of a
pressure-induced valence transition is revealed in YbIr$_2$Si$_2$.

High quality single crystals of YbIr$_2$Si$_2$ have been grown
from In flux in closed Ta crucibles \cite{Hossain 2005}. Depending
on the synthesis conditions, YbIr$_2$Si$_2$ can crystallize either
in the I-type ThCr$_2$Si$_2$ (I4/mmm) as in YbRh$_2$Si$_2$ or in
the P-type CaBe$_2$Ge$_2$ (P4/nmm) structure. The P-type
YbIr$_2$Si$_2$ is magnetically ordered below 0.7 K, whereas the
I-type is a paramagnet \cite{Hossain 2005}. Here we investigate
the properties of the I-type YbIr$_2$Si$_2$. High sensitivity, AC
four-point measurements of the electrical resistivity under high
pressure were carried out in a miniature Bridgman cell ($p\leq 10
~{\rm GPa}$) and a piston-cylinder cell ($p<3 ~{\rm GPa}$), filled
with either steatite (the former) or fluorinert (the latter) as
pressure medium \cite{Yuan 2003a}. The pressure is determined from
the superconducting transition temperature of Pb (or Sn) mounted
inside the cell together with the samples. The experiments were
carried out in a PPMS (down to 1.8 K), a home-made adiabatic
demagnetization cooler ($\sim 250$ mK) and a commercial dilution
refrigerator ($\sim 50$ mK).

\begin{figure}
\centering
\includegraphics[width=0.8\columnwidth]{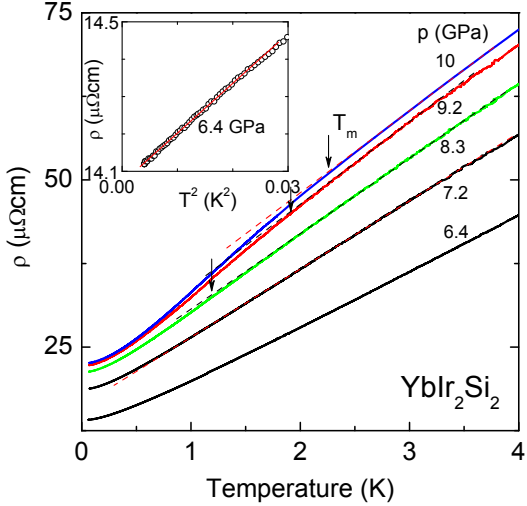}
\caption{(Color online) The electrical resistivity $\rho(T)$
($I\parallel ab$) for YbIr$_2$Si$_2$ at various pressures. The
downward curvature in $\rho(T)$ as marked by the arrows at $T_m$
indicates the occurrence of a magnetic transition above 8.3 GPa.
Inset: The resistivity $\rho(T)$ plotted as a function of $T^2$ at
$p=6.4$ GPa.}
\end{figure}

Fig. 2 shows $\rho(T)$ ($I\parallel ab$) of YbIr$_2$Si$_2$ at
various pressures. At $p=0$, the residual resistivity $\rho_0$ of
this sample is about 5$\mu\Omega$cm (RRR$\simeq 35$). At $p<8$
GPa, YbIr$_2$Si$_2$ is a paramagnet showing a LFL ground state
with $\Delta\rho(T)=\rho(T)-\rho_0=AT^2$ at $T\leq T_{FL}$
($T_{FL}=150\sim200$ mK) (see inset). With increasing temperature,
the resistivity deviates from LFL behavior, following
$\Delta\rho(T)\sim T^\alpha$ with $\alpha\sim 1$. It is noted that
no evidence of magnetism and superconductivity has been detected
down to 50 mK even in highly pure samples ($\rho_0\sim 0.3$
$\mu\Omega$cm, RRR $\simeq 350$) measured in a hydrostatic
piston-cylinder cell. For $p> 8$ GPa, a weak kink at $T_m$ as
marked by the arrows in Fig. 2 is observed in the resistivity
$\rho(T)$, which becomes more pronounced with increasing pressure.
This transition closely resembles the magnetic transitions as
found in other Yb-compounds \cite{Trovarelli 2000, Plessel 2003,
Knebel 2005, Alami 1998}, suggesting a magnetic nature of the
transition at $T_m$. In this context, $T_m$ is determined as the
temperature below which $\rho(T)$ shows a downward deviation from
the $T$-linear resistivity as indicated by the dashed lines in
Fig. 2.

\begin{figure}
\centering
\includegraphics[width=0.9\columnwidth]{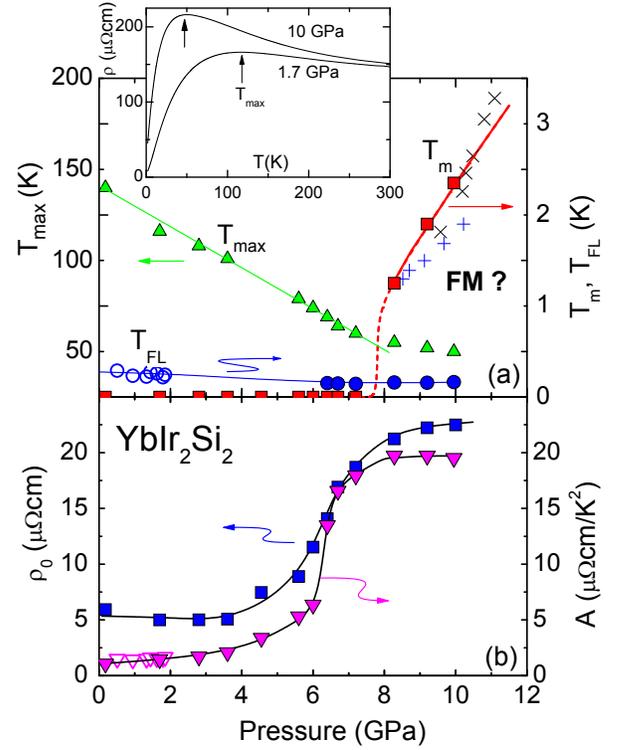}
\caption{(Color online) Pressure dependence of (a) $T_m$, $T_{FL}$
and $T_{max}$, and (b) the residual resistivity $\rho_0$ and the
resistivity coefficient $A$ for YbIr$_2$Si$_2$. The lines are used
as guidance for the eyes. The inset plots $\rho(T)$ over the whole
temperature range for $p=1.7$ GPa and 10 GPa. $T_m(p)$ of
YbRh$_2$Si$_2$ ($\times$, the high-$p$ phase, from Ref.
\cite{Knebel 2005}) and of YbCu$_2$Si$_2$ (+, from Ref.
\cite{Alami 1998}) are included for comparison. The open symbols
represent data from the sample with RRR $\simeq 350$ obtained in a
clamped cell and the filled symbols are for a sample with RRR
$\simeq 35$ measured in a Bridgman cell. No significant difference
can be seen in either $T_{FL}$ or $A$ for these two samples.}
\end{figure}

In Fig. 3a, the derived values for $T_m$ and $T_{FL}$ are plotted
as a function of pressure for YbIr$_2$Si$_2$. Also included in the
figure is the value of $T_{max}$ at which $\rho(T)$ reaches a
maximum attributed to the onset of coherent Kondo scattering. The
value of $T_{max}$ usually scales with the Kondo temperature
$T_K$. In contrast to the Ce- based HF systems, $T_{max}$ of
YbIr$_2$Si$_2$ monotonically decreases with increasing pressure
($dT_{max}/dp=-11.3$ K/GPa), becoming saturated at $p_c\simeq 8$
GPa above which a magnetic transition appears. Similar features of
$T_{max}(p)$ were observed also in YbCu$_2$Si$_2$ \cite{Alami
1998}, but not in YbRh$_2$Si$_2$ \cite{Trovarelli 2000, Plessel
2003} and YbNi$_2$Ge$_2$ \cite{Knebel 2001}. In the latter two
compounds, the resistivity maximum at $T_{max}$ is split into two
maxima under pressure, corresponding to the contributions from the
Kondo effect and the crystalline-electric-field (CEF) as
frequently observed in the Ce-based HF compounds.

For comparison, the magnetic transitions $T_m(p)$ of
YbRh$_2$Si$_2$ (the high-pressure phase only) \cite{Plessel 2003}
and YbCu$_2$Si$_2$ \cite{Alami 1998} are included in Fig. 3(a). In
all these compounds, a magnetic transition appears to abruptly
develop above a certain critical pressure ($p_c\sim 8$ GPa),
showing a uniform magnetic phase diagram and suggesting a
first-order QPT at $p_c$. Note that one can not completely exclude
the possibility that the weak magnetic transition is smeared out
by the enhanced residual scattering while approaching $p_c$. The
magnetic properties of these different Yb compounds appear to be
rather similar at sufficiently high pressures, and one may
speculate about the nature of the magnetic transition at $T_m$
from the following experimental facts: (i) In both YbIr$_2$Si$_2$
\cite{Hossain 2005} and YbRh$_2$Si$_2$ \cite{Gegenwart 2005}, the
Sommerfeld-Wilson ratio is strongly enhanced compared to other HF
systems, indicating that these Yb-compounds are close to a FM
instability. (ii) The NMR results demonstrated that AFM
fluctuations compete with FM fluctuations in YbRh$_2$Si$_2$
\cite{Ishida 2002}. Upon applying pressure, FM fluctuations may
dominate and, therefore, favor a FM-type magnetic structure. (iii)
In YbRh$_2$Si$_2$, the $^{170}$Yb-M\"{o}ssbauer effect measurement
suggests that the large magnetic moments in the high-pressure
phase are aligned along the $\it c$-axis \cite{Plessel 2003}. All
these features indicate that the pressure-induced transition at
$T_m$ is likely of FM nature even though other possibilities can
not be totally excluded at this moment. Further experiments, e.g.,
neutron scattering, are required to confirm its true nature.

In the NFFL model \cite{Pfleiderer 1997, Moriya 1985,
Hertz-Millis}, it is predicted that the conventional LFL behavior
survives well below a crossover temperature $T^*$ in both the
paramagnetic state and the FM state. Above $T^*$, the NFFL model
reduces to a marginal Fermi-liquid model which allows one to
explain the non-Fermi-liquid (NFL) behavior observed at
$T>T_{FL}$. $T^*$ usually vanishes as a magnetic QCP is
approached: $T^*$ is inversely proportional to the magnetic
correlation length $\xi$ which becomes divergent at a QCP.
However, in the case of a first-order magnetic transition near
$p_c$, $T^*$ will be finite \cite{Pfleiderer 1997}. The
experimentally observed FL behavior in the pressurized
YbIr$_2$Si$_2$ is in fact consistent with the NFFL model, further
supporting a FM-type transition at $T_m$.

Fig. 3b shows, for YbIr$_2$Si$_2$, the pressure dependence of
$\rho_0$ and the resistivity coefficient $A$. One notes that the
base temperature for data collected in the Bridgman cell is down
to $250$ mK for $p\leq$6 GPa and $50$ mK for $p>6$ GPa. The fit of
$\Delta \rho=AT^2$ over 0.25K $\leq T\leq0.5$K can give a
reasonable estimation to the coefficient $A$ for $p\leq 6$ GPa, as
evidenced by the good agreement with data at lower temperatures
measured either in the piston-cylinder cell ($p<2$ GPa) or in the
Bridgman cell at higher pressures ($p>6$ GPa). An important
feature here is the steep increase of both $\rho_0$ and $A$
setting in, upon increasing pressure, around 6 GPa. Similar
phenomena were also observed in YbCu$_2$Si$_2$ \cite{Alami 1998}
and MnSi \cite{Thessieu 1995}, the latter of which was regarded as
a typical example to study the FM QPT. Since the characters of the
quasiparticles, e.g., the FL ground state with a nearly
pressure-independent $T_{FL}$, appears to be hardly affected while
crossing $p_c$, it is unlikely that here the enhancement of both
$\rho_0$ and $A$ is mainly due to spin fluctuations as usually
discussed for systems at a magnetic QCP. Note that $T_{FL}$
dramatically increases as tuning away from the QCP in MnSi
\cite{Pfleiderer 1997}. Alternatively, the occurrence of these
unique features in YbIr$_2$Si$_2$ might be related to a valence
change as discussed below.

\begin{figure} \centering
\includegraphics[width=0.86\columnwidth]{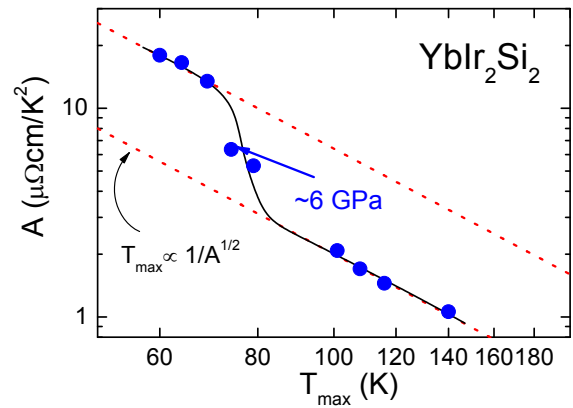}
\caption{(Color online) The log-log plot of the coefficient $A$
and $T_{max}$ (with $p$ as an implicit parameter), showing that
$AT_{max}^2=const.$ with distinct constants for $p> 6$ GPa and
$p<6$ GPa, respectively. }
\end{figure}

In most cases, intermediate-valence compounds have a much lower KW
ratio $A/\gamma^2$ ($\sim 0.4$ $\mu\Omega$cm mol$^2$ K$^2$
J$^{-2}$) than the HF compounds (10 $\mu\Omega$cm mol$^2$ K$^2$
J$^{-2}$) due to the full degeneracy of the ground state
\cite{Tsujii 2005}. In YbIr$_2$Si$_2$ and YbCu$_2$Si$_2$, the CEF
splitting $\Delta_{CEF}$ can be comparable to the Kondo
temperature, leading to a moderate value of the KW ratio:
$A/\gamma^2=2-5$ $\mu\Omega$cm mol$^2$ K$^2$ J$^{-2}$. Upon
applying pressure, $T_K$ is reduced and the ground state could be
strongly affected by the CEF effect. As a result, the KW ratio can
be enhanced due to the reduction of the $f$-orbital degeneracy.
Since $T_{max}\sim T_K \sim \gamma^{-1}$, the value of
$A/\gamma^2$ can be measured by $AT_{max}^2$. In Fig.4, we plot
$A$ vs. $T_{max}$ in a log-log plot, in which the dashed lines
follow $A\sim T_{max}^{-2}$. One can see that a transition takes
place around $p=6$ GPa, on either side of which $AT_{max}^2$ is a
constant, but with different values. For 6 GPa $<p<8$ GPa, a large
value of KW ratio ($A/\gamma^2\sim 20\mu\Omega$cm mol$^2$ K$^2$
J$^{-2}$) is derived from the scaling of $AT_{max}^2$. The
decrease of $A/\gamma^2$ around 6 GPa indicates a weak valence
transition attributed to the change of $f$-orbital degeneracy.
This assertion is further supported by the the resistivity
isotherms $\Delta\rho(T,p)=\rho(T,p)-\rho_0(p)$ as shown in Fig.
5. The large decrease of $\Delta\rho(T,p)$ (at low-$T$) and $A$
below $p=6.4$ GPa implies a weakening of electronic correlations
as a result of the delocalization of $f$-electrons at lower
pressures.

\begin{figure}
\centering
\includegraphics[width=0.90\columnwidth]{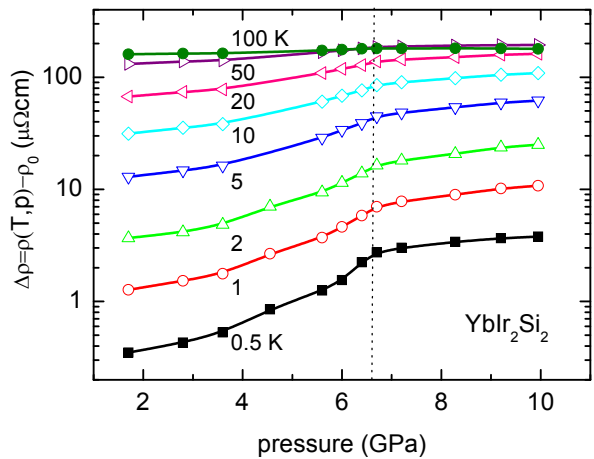}
\caption{(Color online) Pressure dependence of the resistivity
isotherms $\Delta\rho(p)$ ($=\rho(T,p)-\rho_0(p)$) at various
temperatures.}
\end{figure}

Similar features of $\Delta\rho(T,p)$, $\rho_0$ and the
coefficient $A$ as demonstrated in YbIr$_2$Si$_2$ were previously
found in CeCu$_2$(Si$_{1-x}$Ge$_x$)$_2$ at a weak first-order
volume-collapse transition \cite{Yuan 2003}. Indeed, a theoretical
model based on valence fluctuations predicts a pronounced
enhancement of $\rho_0$ and $A$ as well as the $T$-linear
resistivity above a crossover temperature $T_v$ at a valence
transition \cite{Miyake 2002}, consistent with our findings here.
How these two (magnetic and valence) transitions are interacting
with each other remains an interesting question which can not,
however, be answered based on the available results. For example,
in YbCu$_2$Si$_2$ the magnetic transition appears around 8 GPa,
but $\rho_0$ and $A$ peak at 10 GPa, where the KW relation starts
to deviate \cite{Alami 1998}. On the other hand, in YbIr$_2$Si$_2$
the valence transition appears prior to the magnetic transition.
Interestingly, the valence transition seems to be accompanied by a
strong enhancement of $\rho_0$ and $A$.

The cause for the unique appearance of the low-pressure AFM phase
in YbRh$_2$Si$_2$ \cite{Trovarelli 2000, Plessel 2003, Knebel
2005} (and possibly also in YbNi$_2$Ge$_2$ \cite{Knebel 2001})
remains unclear. One possibility might be related to the band
structure, e.g., the distinct values of the density of state (DOS)
at the Fermi energy $N(E_F)$, originating from the different
electronic configurations of Rh and Ir.  While the intrasite
coupling constant $J$ between $f$- and conduction- electrons may
experience similar modulation under pressure, the value of
$N(E_F)J$ may vary from compound to compound because of the
different values of $N(E_F)$. In YbRh$_2$Si$_2$, $N(E_F)$ can be
small because the DOS peaks just below $E_F$ \cite{Wigger 2006}.
The resulting small value of $N(E_F)J$ then may favor magnetic
ordering even at low pressures, while in YbIr$_2$Si$_2$ and
YbCu$_2$Si$_2$ the Kondo effect may dominate in the same pressure
region, leading to a nonmagnetic ground state \cite{Doniach 1977}.
These arguments are compatible with the pressure dependence of
$T_{max}$. $T_{max}$ reaches a few kelvins in YbRh$_2$Si$_2$
\cite{Plessel 2003} and YbNi$_2$Ge$_2$ \cite{Knebel 2001}, but 50
K in YbIr$_2$Si$_2$ and YbCu$_2$Si$_2$ \cite{Alami 1998} at $p_c$,
indicating a predominant RKKY interaction in  YbRh$_2$Si$_2$. The
intersite RKKY coupling can change sign and, therefore, give rise
to different type of magnetic ordering under pressure.

In summary, we have shown that a FM-type quantum phase transition
is likely to exist in the Yb-compounds under sufficiently high
pressure and that the related properties may be well described
within the NFFL model. These findings will be essentially
important for understanding the unusual properties of
YbRh$_2$Si$_2$, especially its complex $p-T$ phase diagram, and
will stimulate further exploration of FM quantum criticality in
the HF systems. Moreover, a weak valence transition accompanied by
a huge enhancement of $\rho_0$ and the resistivity coefficient $A$
appears to exist in these pressurized Yb-systems. To elucidate
these remarkable properties, further experimental and theoretical
efforts are highly desired.

We thank Q. Si and M. B. Salamon for stimulating discussions. HQY
acknowledges the ICAM fellowship.


\begin{thebibliography}{apssamp}

\bibitem{Thalmeier 2004} See, e.g., P. Thalmeier {\it et al.},
cond-mat/0409363, (2005).

\bibitem{Custers 2003} J. Custers {\it et al.}, Nature {\bf 424},
524 (2003).

\bibitem{Thessieu 1995} C. Thessieu {\it et al.}, Solid State
Commun. {\bf 95}, 707 (1995).

\bibitem{Pfleiderer 1997} C. Pfleiderer {\it et al.}, Phys. Rev. B {\bf 55}, 8330 (1997); N.
Doiron-Leyraud {\it et al}, Nature {\bf 425}, 595 (2003).

\bibitem{Grigera 2001} S. A. Grigera {\it et al.}, Science {\bf
294} 329 (2001); S. A. Grigera {\it et al.}, Science {\bf 306}
1154 (2004); P. Gegenwart {\it et al.}, Phys. Rev. Lett. {\bf 96},
136402 (2006).

\bibitem{Yuan 2003} H. Q. Yuan {\it et al.}, Science {\bf 302}, 2104 (2003);
H. Q. Yuan {\it et al}, Phys. Rev. Lett. {\bf 96}, 047008 (2006).

\bibitem{Trovarelli 2000} O. Trovarelli {\it et al.}, Phys. Rev.
Lett. {\bf 85}, 626 (2000).

\bibitem{Gegenwart 2002} P. Gegenwart {\it et al.}, Phys. Rev.
Lett. {\bf 89}, 056402 (2002).

\bibitem{Plessel 2003} J. Plessel {\it et al.}, Phys. Rev. B {\bf
67}, 180403(R) (2003).

\bibitem{Knebel 2005} G. Knebel {\it et al.}, Physica B {\bf
359-361}, 20 (2005).

\bibitem{Ishida 2002} K. Ishida {\it et al.}, Phys. Rev. Lett. {\bf 89},
107202 (2002).

\bibitem{Gegenwart 2005} P. Gegenwart {\it et al.}, Phys. Rev. Lett.
{\bf 94}, 076402 (2005).

\bibitem{Hossain 2005} Z. Hossain {\it et al.}, Phys. Rev. B {\bf
72}, 094411 (2005).

\bibitem{Yuan 2003a} H. Q. Yuan, Ph.D thesis, Technische
Universit\"{a}t Dresden, 2003.

\bibitem{Alami 1998} K. Alami-Yadri {\it et al.},
Eur. Phys. J. B {\bf 6}, 5 (1998); W. Winkelmann {\it et al.},
Phys. Rev. B {\bf 60}, 3324 (1999).

\bibitem{Knebel 2001} G. Knebel {\it et al.}, J. Phys.: Condens.
Matter {\bf 13}, 10935 (2001).

\bibitem{Moriya 1985} T. Moriya, {\it Spin fluctuations in itinerant
electron magnetism} (Springer, Berlin, 1985).

\bibitem{Hertz-Millis} J. A. Hertz, Phys. Rev. B {\bf 14}, 1165
(1976); A. J. Millis, Phys. Rev. B {\bf 48}, 7183 (1993).

\bibitem{Tsujii 2005} N. Tsujii {\it et al.},
Phys. Rev. Lett. {\bf 94}, 057201 (2005).

\bibitem{Miyake 2002} K. Miyake and H. Maebashi, J. Phys. Soc.
Jpn. {\bf 71}, 1007 (2002); A. T. Holmes {\it et al.}, Phys. Rev.
B {\bf 69}, 024508 (2004).

\bibitem{Wigger 2006} G. A. Wigger {\it et al.}, to be published (2006).

\bibitem{Doniach 1977} S. Doniach, Physica B {\bf 91}, 231 (1977).

\end{thebibliography}
\end{document}